# Low Vibration Laboratory with a Single-Stage Vibration Isolation for Microscopy Applications

Bert Voigtländer,[1, 2, *] Peter Coenen,[1, 2] Vasily Cherepanov,[1, 2]
Peter Borgens,[1, 2] Thomas Duden,[3] and F. Stefan Tautz[1, 2]

[1]*Peter Grünberg Institut (PGI-3), Forschungszentrum Jülich, 52425 Jülich, Germany*
[2]*Jülich Aachen Research Alliance (JARA) Fundamentals of Future Information Technology, 52425 Jülich, Germany*
[3]*Konstruktionsbüro Duden, Borgsenallee 35, 33649 Bielefeld, Germany*

The construction and the vibrational performance of a low vibration laboratory for microscopy applications comprising a 100 ton floating foundation supported by passive pneumatic isolators (air springs), which rest themselves on a 200 ton solid base plate is discussed. The optimization of the air spring system lead to a vibration level on the floating floor below that induced by an acceleration of 10 ng for most frequencies. Additional acoustic and electromagnetic isolation is accomplished by a room-in-room concept.

PACS numbers:

## INTRODUCTION

Modern microscopes with a spatial resolution reaching deep into the nanoscale or even into the picometer scale are known to be sensitive to external vibrations. This problem occurs for different kinds of microscopes in different ways. Scanning electron microscopes and transmission electron microscopes are specifically sensitive to horizontal vibrations, while scanning probe microscopy techniques are much more sensitive to vertical vibrations, since picometer resolution is mainly desired in the vertical direction.

A standard approach in scanning probe microscopy is to combine a stiff construction of the microscope itself with a vibration isolation using springs and an eddy current damping. This approach benefits from the fact that not the vibration of the sample itself, but the *difference* between the sample and the tip motions have to be minimized [1].

While scanning probe microscopes are small and can be suspended by springs, this is impossible for most electron microscopes, which have a huge mechanical loop from the sample to the microscope. Under some conditions also for scanning tunneling microscopes a spring suspension cannot be used. For example, at low temperatures a spring suspension prevents good thermal coupling to the microscope or, if large magnetic fields are present, an eddy current damping cannot be used. In these cases either the whole microscope chamber or even the whole laboratory in which the electron microscope or the low temperature scanning probe microscope is operated should be isolated from external vibrations, such as ground vibrations and acoustic vibrations.

In scanning probe microscopy multi-stage vibration isolation stages have been analyzed theoretically [2, 3] and have been implemented and tested in practice [4, 5]. Several multi-stage vibration isolation approaches have been followed and compared in the recent years. For instance, either a concrete block ($\sim 56$ tons) was placed on the base slab of the building [4] or alternatively a concrete block ($\sim 37$ tons) rests on a separate foundation ($\sim 47$ tons) not connected to the base slab of the building [5]. In these approaches a second stage of vibration isolation was implemented by suspending a concrete table ($\sim 8$ tons) on air springs. The actual microscope chamber was suspended in a third stage of vibration isolation using passive isolation dampers [4]. Another example of a three-stage vibration isolation system with a mass of $\sim 56$ tons in the first stage has lead to an excellent performance [6].

In spite of the success of these multi-stage approaches for vibration isolation, we, as well as others [7], decided for a single-stage approach, because (a) multi-stage approaches involve more effort, (b) the subsequent stages beyond the first do not lead to a substantial improvement of the vibration isolation [6], and (c) the one-stage approach results in a general purpose low vibration laboratory which can be used for any kind of equipment, while the multi-stage approaches are specifically tailored to a particular instrument, because stages 2 and 3 have to be engineered into the instrument.

In the following we describe the design of this low vibration laboratory and also mention the pitfalls and problems we encountered, so that these can be prevented in future realizations of similar low vibration laboratories. Our single-stage design provides vibration isolation of SPM instruments comparable to that of the best three-stage systems.

## VIBRATION ISOLATION BASICS

The fundamental concept in vibration isolation is to use the isolation properties of a damped harmonic oscillator, which is characterized by its resonance frequency $\omega_0$ and damping. The damping is characterized by the quality factor of the oscillator $Q$. The damped oscillator is considered to be driven by external noise oscillations of frequency $\omega$. The transfer function of a driven damped harmonic oscillator is the ratio of the oscillation amplitude to the excitation amplitude (absolute values), and can be written as [1]

$$\kappa(\omega) = \sqrt{\frac{1 + \frac{1}{Q^2}\left(\frac{\omega}{\omega_0}\right)^2}{\left(1 - \left(\frac{\omega}{\omega_0}\right)^2\right)^2 + \frac{1}{Q^2}\left(\frac{\omega}{\omega_0}\right)^2}}. \qquad (1)$$

In Fig. 1 the transfer function for a driven damped harmonic oscillator with a resonance frequency $f_0 = \omega_0/(2\pi) = 0.8\,\mathrm{Hz}$ is shown for three different quality factors $Q = 2, 5$, and 100. The vibration isolation properties of a harmonic oscillator can be described as follows. Below the resonance frequency $\omega_0$ vibrations are transmitted with a transfer function close to one. In the vicinity of the resonance frequency the transfer function has a value close to $Q$, which means that (noise) vibrations are amplified $Q$ times. An actual vibration isolation occurs only above the resonance frequency. For high quality factors the transfer function decreases as $\kappa \propto 1/\omega^2$. This results in a decrease of the transfer function by a factor of 100 (40 dB) per decade frequency increase. This excellent damping leads on the other hand to an enormous resonance amplitude (the black curve in Fig. 1 corresponds to $Q = 100$). In contrast, for small quality factors (the green curve in Fig. 1 corresponds to $Q = 2$) the damping behavior above the resonance is less effective, approaching $\kappa \propto 1/\omega$ for large frequencies and for small $Q$-factors, corresponding to a decrease of the transfer function by only a factor of 10 (20 dB) per decade frequency increase. In practice, a compromise has to be found between good damping properties beyond the resonance frequency of the oscillator and a small overshoot at the resonance.

The experimentally measured transfer function for vertical oscillations of the final configuration of our single-stage vibration isolation system, which we describe in detail later, is shown in red in Fig. 1. The transfer function was measured using the naturally present ground vibrations as excitation, i.e. without intentionally exciting (shaking) the suspended laboratory mass. The measured transfer function follows the behavior of a harmonic oscillator with a resonance frequency of $f_0 = 0.8\,\mathrm{Hz}$ and a quality factor $Q = 5$ up to a frequency of 3-4 Hz. For larger frequencies acoustic effects (which will be discussed below) come into play, leading to a damping behavior worse than that expected for the harmonic oscillator model.

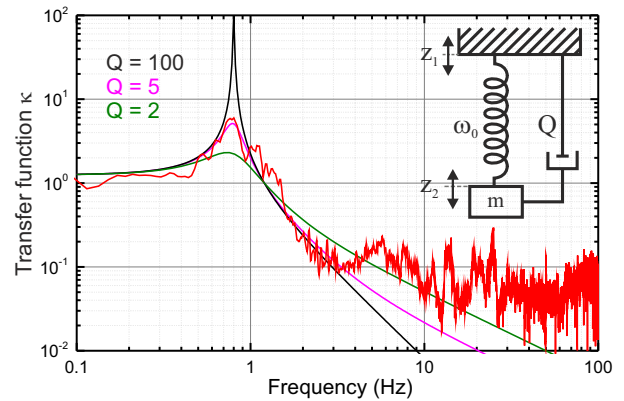

FIG. 1: (color online) Transfer function of a driven damped harmonic oscillator with a resonance frequency of 0.8 Hz, displayed for three different $Q$-factors. Additionally, the data (red) show the measured transfer function for vertical oscillations of an oscillator comprising a $\sim 100$ ton inertial mass (floating floor of the low vibration laboratory) and pneumatic isolators, which act as (air) springs.

## LOW VIBRATION LABORATORY

### Design of the low vibration laboratory

The laboratory with two low vibration measurement cabins was built in an existing hall, as shown in Fig. 2. The cabins are resting on a concrete foundation of a mass of $\sim 200$ tons with dimensions of $14 \times 6 \times 0.65\,\mathrm{m}^3$. The floating foundations for each chamber have a mass of $\sim 100$ tons each and are supported by four passive pneumatic isolators (CFM-Schiller type GRB 2480 ZV) [8]. The resonance frequency of this configuration would be 1.2 Hz. Supplementary air volumes of 300 liter were connected to each air spring in order to lower the resonance frequency to 0.6 Hz. The mass of the floating foundation was increased by adding lead bricks into the concrete foundation. The vibration isolation properties of this single-stage vibration isolation will be presented in detail below.

Apart from the isolation against ground vibrations, also acoustic noise can transmit to the sensitive equipment and induce horizontal as well as vertical vibrations. Our approach to shield the laboratory from acoustic excitations is the room-in-room concept. An outer brick wall (Fig. 2) rests on the large foundation, while an inner drywall rests on the suspended cabin foundation. The hard outer brick wall shell protects the inside of the low vibration laboratory against acoustic noise.

In our case, due to height limitations of the hall in which the low vibration laboratory was constructed, the ceiling consists of wooden planks covered by flagstones. For optimized acoustic shielding it would have been better to implement a solid (concrete) ceiling in order to have a hard outer shell protecting against acoustic noise.

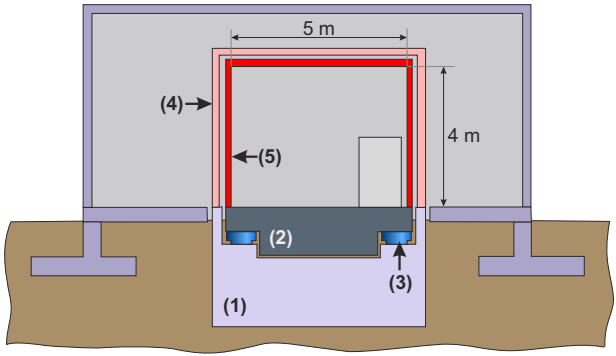

FIG. 2: (color online) The low vibration laboratory was built in an existing hall (gray). The measurement cabins rest on a foundation of a mass of 200 tons (1). The single-stage vibration isolation insulating the cabin against ground vibrations consists of the floating foundation (2) of a mass of $\sim 100$ tons and four passive pneumatic isolators (3), acting as springs. Acoustic isolation is provided by a hard brick wall enclosure (4) resting on the large foundation, and an inner drywall (5) resting on the suspended cabin foundation.

The inside of the laboratory was completely covered by an electromagnetic shielding (Series 81 from ETS Lindgren [9]). In one of the two laboratories the shielding was made from magnetic material, providing better shielding values, while in the other laboratory the electromagnetic shielding is realized using non-magnetic materials, allowing for conditions of very high magnetic fields in this laboratory.

The electrical power to the laboratory is provided through the entrance door for maintenance purposes, while under measurement conditions the sound- and electromagnetic-proof door is closed and the cabling to the measurement equipment is guided through dedicated panels, in which the cables can be vibrationally shielded and high frequency shielded, if required.

**Vibration measurement**

Vibrations are measured with piezoelectric transducers which measure the corresponding acceleration. In our case we use 731A seismic accelerometers and a P31 power unit, both from Wilcoxon Research. The accelerometers were calibrated by the supplier company, and we refer to the factory calibration. The spectra are acquired with a 16 bit USB oscilloscope, at a sampling rate of 40 k samples/s, with a low pass filter set to 100 Hz. About 2 M samples are taken in 50 seconds, and subsequently Fourier transformed. Ten spectra are averaged to deliver the required sensitivity for the results presented in Figs. 3, 5 and 6. Considering a sinusoidal vibration $z = z_0 \cos(\omega t)$, the velocity and acceleration are written

as

$$v = \dot{z} = -z_0\omega \sin(\omega t) := -v_0 \sin(\omega t) \qquad (2)$$
$$a = \ddot{z} = -z_0\omega^2 \cos(\omega t). \qquad (3)$$

These relations can be used in order to convert between oscillation amplitude, velocity and acceleration. The corresponding rms quantities, e.g. of the vibration velocity, relate to the amplitude $v_0$ as $v_{\text{rms}} = \sqrt{\langle v^2 \rangle} = v_0/\sqrt{2}$. The oscilloscope software is set to deliver the spectral density in $\mathrm{dB}V_{\text{rms}}/\sqrt{\mathrm{Hz}}$. The data presented here is at first converted to $\mathrm{dB\,mg}_{\text{rms}}/\sqrt{\mathrm{Hz}}$ using the conversion factor given by the manufacturer, relating the signal to fractions of the gravitational acceleration, and then to velocity by using Eqs.(2) and (3). In this way, the data allow the comparison with other measurements, independent of e.g. sampling details.

The quantity measured by the accelerometers is the spectral density of the acceleration. This is converted to the usually displayed spectral density of the velocity $N_v(f)$ by division through $\omega$, according to the above equations. The spectral density of the velocity $N_v(f)$ in $\mathrm{m/s}/\sqrt{\mathrm{Hz}}$ is related to the power spectral density of the velocity $N_v^2(f)$ by $N_v(f) = \sqrt{N_v^2(f)}$. Both are functions of the natural frequency $f = \omega/(2\pi)$. An important property of the power spectral density of the velocity is that it relates to the mean square of the velocity as

$$\langle v^2 \rangle = \int_0^{\infty} N_v^2(f) df. \qquad (4)$$

If the power spectral density of the velocity is considered within a certain frequency bandwidth $B = f_2 - f_1$ between $f_1$ and $f_2$, the mean square velocity can be written as

$$\langle v^2 \rangle_B = \int_{f_1}^{f_2} N_v^2(f) df. \qquad (5)$$

If the power spectral density is constant between $f_1$ and $f_2$ (white noise), Eq.(5) reduces to

$$\langle v^2 \rangle_B = (f_2 - f_1) N_v^2, \qquad (6)$$

and we obtain

$$v_{\text{RMS}} = \sqrt{\langle v^2 \rangle_B} = N_v \sqrt{B}. \qquad (7)$$

The above equation means that the spectral density of the velocity $N_v$ can be considered as the rms velocity $\sqrt{\langle v^2 \rangle}$ referred to a measurement bandwidth of 1 Hz, assuming a constant spectral density within this range. For example, a spectral density of the velocity of $0.1\,\mu\mathrm{m/s}/\sqrt{\mathrm{Hz}}$ at 20 Hz corresponds to a vibration velocity of $0.1\,\mu\mathrm{m/s}$ arising from a frequency window from 19.5 Hz to 20.5 Hz. Therefore, the spectral velocity density is often also expressed as velocity in units $\mathrm{m/s}/\sqrt{\mathrm{Hz}}$ with the implicit convention that these data are referred to a bandwidth of one Hertz [4–6].

In Figs. 3, 5 and 6 we compare experimental data to green reference lines corresponding to a spectral velocity density calculated from a constant spectral density of the acceleration $N_a$, which we abbreviate as a fraction of the gravitational acceleration $g$. For instance, the label $1\,\mu g$ in Fig. 3 means that $N_a = 9.81 \cdot 10^{-6}/\sqrt{2}\,\mathrm{m/s^2}/\sqrt{\mathrm{Hz}}$. The corresponding spectral density of the velocity (rms) results as $N_v(f) = 1/(2\pi f) \cdot 9.81 \cdot 10^{-6}/\sqrt{2}\,\mathrm{m/s^2}/\sqrt{\mathrm{Hz}}$, using Eq.(2) and Eq.(3) in order to convert from acceleration to velocity. In this paper, our metric of choice is the spectral density of the velocity. A one-third octave band analysis of the final results can be found in the appendix.

While we have considered the transfer function in order to characterize the efficiency of a vibration isolation system, finally however, the actual vibration level at the unit to be isolated from vibrations is the relevant quantity. This vibration level depends also on the level of ground vibrations as the input quantity to the vibration isolation system ($z_1$ in Fig. 1). The vibrational amplitude on the isolated platform is given by the ground vibration amplitude times the transfer function $\kappa$ of the vibration isolation as

$$v_{\mathrm{isolated}}(f) = v_{\mathrm{ground}}(f) \cdot \kappa(f). \qquad (8)$$

Thus in frequency regions in which the vibration level is low from the beginning (low $v_{\mathrm{ground}}(f)$) no vibration isolation is required.

### Initial situation and solid foundation (200 tons)

The spectral density of the ground vibration velocity present in the hall prior to the construction of the vibration isolation system is shown as magenta curve in Fig. 3. For low frequencies it has a level of $0.7\,\mu\mathrm{m/s}/\sqrt{\mathrm{Hz}}$ and drops below $0.1\,\mu\mathrm{m/s}/\sqrt{\mathrm{Hz}}$ beyond 15 Hz. Thus, the vibration level in this hall corresponds about to $1\,\mu g$, as shown by the uppermost green curve. The blue curve in Fig. 3 is measured on the 200 tons concrete block introduced below the hall ((1) in Fig. 2). Note that this measurement was performed on the final system, i.e. with the 100 tons floating foundation already installed and floating. The vibration level on the solid foundation is substantially reduced compared to that of the floor of the hall. Below 20 Hz the improvement corresponds to a factor of 10, i.e from $1\,\mu g$ to 100 ng, while from 70 Hz the improvement approaches a factor of 100, i.e. to 10 ng. The largest peak is located around 25 Hz and is present before and (reduced in amplitude) after the construction of the solid 200 tons foundation. It arises from compressors used in a nearby room for the generation of com-pressed air. Moreover, on the solid foundation a small hump of unknown origin was measured at 10 Hz.

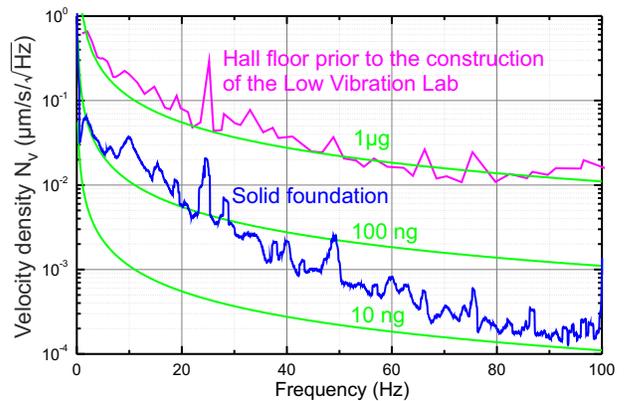

FIG. 3: (color online) Spectral density of the rms vibration velocity (vertical component) $N_v(f)$ on the floor of the building before any construction has been started (magenta line), compared to the rms vibration level on the solid 200 ton solid foundation (blue line). For comparison $N_v(f)$ curves (rms) corresponding to different fractions of the gravitational acceleration $g$ are shown.

In summary, the solid 200 tons concrete foundation serves as an efficient first step in the reduction of the ground vibrations. It reduces the vibration level already by a factor of 10 in the most crucial range below 20 Hz and up to a factor of 100 for larger frequencies.

### Floating foundation: Optimization of the air spring system

The passive pneumatic air spring system consists of four air springs (CFM Schiller type GRB 2480 ZV indicated in Fig. 2 as (3) [8]), as well as an additional volume $V_{\mathrm{add}}$ connected to each air spring via a hose of length $L$ and cross section $A$, as schematically shown in Fig. 4(b).

The initial spectral density of the velocity measured on top of the floating 100 ton foundation is shown in Fig. 5 in blue. The peak at 0.6 Hz corresponds to the fundamental vertical resonance frequency of the oscillator formed by the 100 ton inertial mass and the air springs. A second peak in the vibration spectrum appears slightly above 1 Hz and arises most likely from the specified resonance of the passive pneumatic isolators without the additional volumes [10]. A third, initially unexpected very high and broad peak in the vibrational spectrum arises around 3 Hz. The origin of this peak was identified to be a Helmholtz resonance [11].

The Helmholtz resonance is the phenomenon of an oscillation of an air volume at an opening of a cavity. A generic Helmholtz resonator (Fig. 4(a)) comprises a cavity with compressible gas inside, acting as a spring, and the gas in the neck of the cavity, acting as an oscillating mass. In daily life, Helmholtz resonances are known from the low-key tones beverage bottles can generate when properly blown. Our pneumatic air spring system represents a two-spring Helmholtz resonator configuration, consisting of the original volume of the pneumatic isolators ($V_1 = 0.08 \, \mathrm{m}^3$) and an additional, even larger, volume in a pressure vessel close to the pneumatic isolators ($V_{\mathrm{add}} = 0.3 \, \mathrm{m}^3$). Both volumes are connected by a hose, as shown in Fig. 4(b). Thus, the air spring system with the additional volume represents an oscillator system which is excited by ground vibrations coupling to the volume of the air spring. The frequency of the Helmholtz resonance [11] for the configuration with two volumes is given by the following equation, in which the spring forces of two air volumes connected to a hose junction are combined:

$$f_0 = \frac{c}{2\pi} \sqrt{\frac{A}{L + \Delta L} \left( \frac{1}{V} + \frac{1}{V_{\mathrm{add}}} \right)}, \qquad (9)$$

with $c = 343 \, \mathrm{m/s}$ being the velocity of sound, $A$ the cross sectional area of the connecting hose between the two volumes, $L$ the length of the hose, and $V$ and $V_{\mathrm{add}}$ the volumes of the air springs and the additional volume, respectively. $\Delta L$ is a correction for the orifice, which we set to $\Delta L = 0.035 \, \mathrm{m}$, being approximately twice the diameter of the hose [11]. For a hose diameter of $d = 0.019 \, \mathrm{m}$, and $L = 1.25 \, \mathrm{m}$ a frequency $f = 3.2 \, \mathrm{Hz}$ results, in accord with the experiment.

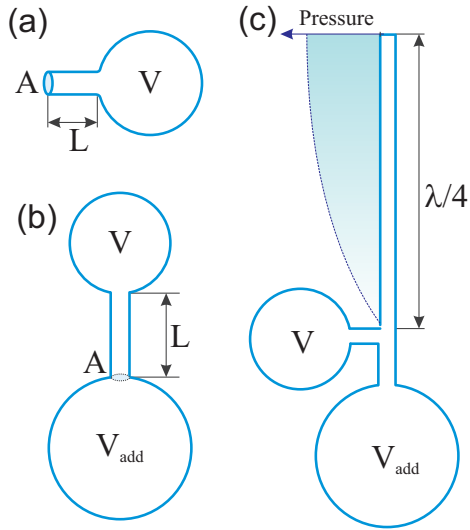

FIG. 4: (color online) (a) Generic Helmholtz resonator. (b) Schematics of the air spring system with the two air volumes and the connecting hose forming a modified two-spring Helmholtz resonator. (c) A hose with a length of a quarter of a wavelength of the sound (at the Helmholtz resonance) is connected to the air spring system. As this hose is closed at its end, destructive interference between the air oscillations in the two-spring Helmholtz resonator and the sound wave reflected at the end of the $\lambda/4$-hose suppresses the Helmholtz resonance.

According to Eq. 9 the Helmholtz resonance shifts up in frequency if the hose diameter $d$ is increased or the hose length $L$ is shortened. In a first attempt to get rid of this undesired resonance frequency the length of the hose was reduced to $L = 0.5 \, \mathrm{m}$. The experimentally measured Helmholtz resonance shifted correspondingly to about 5 Hz and simultaneously the spectral density of the velocity decreased by a factor of ten (gray trace in Fig. 5). This observed shift of the resonance is in accord with the value of 5 Hz expected form Eq. 9 and thus proves the origin of this peak as a Helmholtz resonance. In the final configuration of the vibration isolation system (see below) two hoses with $d = 25 \, \mathrm{mm}$ and lengths $L = 0.75 \, \mathrm{m}$ were used, shifting the Helmholtz resonance according to Eq. 9 to 7.7 Hz, while the resulting resonance peak was measured at 7.4 Hz (not shown).

We have applied two methods to suppress the Helmholtz resonance in our isolation system. The first method aimed at selectively damping the Helmholtz resonance. A hose with a length of one quarter of the wavelength at the frequency of the Helmholtz resonance was connected via the side port of a T-fitting, close to the additional volume (Fig. 4(c)). This hose was closed at the end, giving rise to the formation of a pressure antinode (corresponding to a displacement node) at the closed end of the hose. Since the length of the hose is a quarter of the wavelength of sound at the Helmholtz resonance frequency, a pressure node arises at the other end of the hose, i.e. at the T-fitting. Since the wavelength of the sound wave (70 m) is much larger than the dimension of the air spring system including the additional volume ($< 2 \, \mathrm{m}$), the standing wave in the $\lambda/4$-hose gives rise to a pressure node in the whole air spring system. In this way the excitation of the Helmholtz resonance oscillations is suppressed by destructive interference selectively at the resonance frequency. After tuning the length of the $\lambda/4$-hose, we could measure a clear suppression of vibrations at the position of the former Helmholtz peak. However, the suppression was very sharp, and two side bands remained. The width of the suppression of the Helmholtz resonance was increased by damping the gas oscillation with an energy-absorbing component. To this end, we filled the end of the $\lambda/4$-hose with copper mesh. The copper mesh acts as a thermal absorber and thus equalizes the adiabatic temperature response of the air in the $\lambda/4$-hose close to its closed end (the maximum of pressure). This reduced the resonant response by some dB, rendering the isolation system slightly damped. In a further step, some copper mesh was introduced into the connection between 300 l vessel and pneumatic isolator. These provided additional friction and finally led to the desired essential elimination of the Helmholtz resonance (red trace in Fig. 5), however, at the expense of slightly higher overall spring stiffness, indicated by a shift of the fundamental resonance to a value slightly larger than 1 Hz.

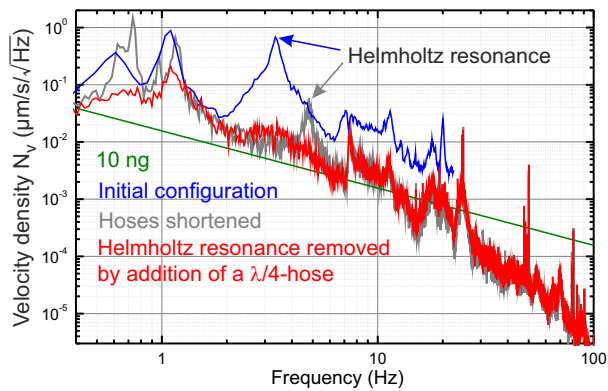

FIG. 5: (color online) Spectral density of the rms vibration velocity (vertical component) for different stages of the optimization of the vibration isolation. The initial state (blue) is characterized by a large peak at about 3 Hz due to the Helmholtz resonance. After shortening of the hoses the Helmholtz resonance shifted to about 5 Hz (gray). The Helmholtz resonance has been removed by destructive interference of standing waves using an additional $\lambda/4$-hose (red).

Alternatively, the Helmholtz resonance can be suppressed by using two larger diameter hoses ($d = 25 \, \mathrm{mm}$) between the two volumes $V$ and $V_{add}$, benefiting from the increase of the frequency of the Helmholtz resonance and the accompanying lower spectral density of the velocity for larger hose diameters according to Eq. 9. Because the frequency difference between the fundamental resonance of the isolator (0.6 Hz) and the Helmholtz resonance (7.4 Hz) is sufficiently large, the amplitude of the Helmholtz resonance became quite small. In order to further dampen the Helmholtz resonance, the hoses were filled by several layers of copper mesh at the junctions to the pneumatic isolators and the 300 l tanks, in order to provide turbulence damping. The damping was effectively suppressing the Helmholtz resonance, which does not appear noticeably in Fig. 6, which shows the results measured on the final arrangement of the isolation system (two hoses, $d = 25 \, \mathrm{mm}$ and length $L = 0.75 \, \mathrm{m}$). Thus the larger effort of the installation of the $\lambda/4$-hoses was avoided. This damping has also the positive effect of reducing the quality factor of the fundamental vertical resonance frequency of the air spring system. The quality factor of the lowest vertical resonance below 1 Hz was in some configurations larger than ten, e.g. in the configuration with shortened hoses shown in Fig. 5. The damping induced by filling the hoses with several layers of copper mesh reduced the quality factor to a value of about five, as seen from Fig. 1.

### Vibrational performance of the optimized system

In a final optimization step the acoustic shielding was optimized by closing the revision openings to the pits in which the pneumatic isolators are located. By this a complete sound protection of the floating foundation was achieved and the sail-effect, by which pressure variations that are induced in the volume between the two walls (4) and (5) in Fig. 2 (by sound in the hall) may move the inner wall (5), was minimized. The acoustic shielding reduced the measured spectral density of the velocity in the frequency ranges between 5 Hz and 20 Hz. However, as an undesired side effect it created a closed volume of air which increased the stiffness of the fundamental resonance, i.e. an increase of the lowest vertical resonance frequency from initially 0.6 Hz to 0.8 Hz in the final configuration. The effect of an increase of the fundamental resonance frequency due to the acoustic enclosure was confirmed by comparative measurements with opened and closed revision openings.

After all, the inclusion of the additional volumes created several follow-up problems which should be avoided by using larger volumes of the pneumatic isolators from the start, if the building conditions allow for the use of large-volume pneumatic isolators. If there is not enough space to allow the use of large isolators, additional volumes can be used and good vibration isolation can be achieved by implementing the measures described above, in particular in order to avoid the Helmholtz resonance.

The spectral density of the velocity measured in the final configuration on top of the floating floor are shown in comparison to the 10 ng line in Fig. 6. The vertical vibration level (shown in red) is in most frequency ranges (apart from the fundamental resonance) at or below the 10 ng line. The enhanced spectral density of the velocity around 25 Hz is also present in the excitation spectrum (solid foundation) and arises form asynchronous motors from the building services. While for scanning probe microscopy applications the vibrational noise in the vertical direction is most important, for SEM and TEM applications the measured horizontal vibration level is shown as gray line in Fig. 6 and has also a similarly low level. Our results can be compared to those published for other labs [4–7].

### CONCLUSIONS

A low vibration laboratory with a vibration level below that of an acceleration of 10 ng was constructed. This vibration level was reached with a single-stage passive vibration isolation avoiding the complexities of a more elaborate multi-stage or active vibration isolation systems. The vibration isolation stage comprises a 200 ton concrete baseplate and a 100 ton floating floor, which rests on passive pneumatic isolators. Additional exter-nal volumes were connected to the pneumatic isolators in order to reduce the lowest resonance frequency to 0.6 Hz. However, the connection of the additional volumes had the side effect of a Helmholtz resonance. We present approaches to eliminate the Helmholtz resonance and to reduce the initially high quality factor of the vibration isolation system. Effective acoustic and electromagnetic shielding of the low vibration laboratory was implemented by a room-in-room concept. The approach described here provides a generic low vibration laboratory, which can be used for any kind of vibrationally sensitive equipment installed inside. It reaches a performance at least as good as more elaborate multi-stage vibration isolation systems.

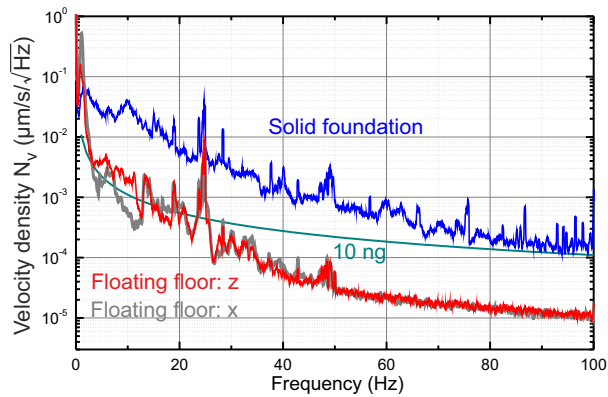

FIG. 6: (color online) Spectral density of the velocity (rms) measured on the floating floor in the final configuration. The vertical direction is shown in red, while one horizontal direction is shown in gray. The vertical spectral density of the velocity lies in most frequency ranges at or below the 10 ng line is shown in green.

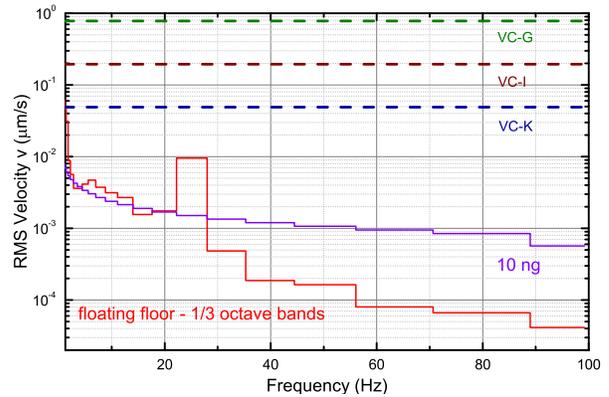

FIG. 7: (color online) One-third octave band analysis of the RMS velocity $v$ in $z$-direction measured on the floating floor in the final configuration. The red curve is calculated from the corresponding curve in Fig. 6. Standards for vibration isolation of buildings (vibration criteria, VC) are given for comparison (following IEST-RP-CC024), as well as the 10 ng line (in terms of one-third octave bands).

## ACKNOWLEDGMENTS

We would like to thank G. Malat and S. Schmücker from the planning and building services of the Forschungszentrum Jülich, as well as B. Fortmeyer and M. Walter from the pbr AG, and S. Tonutti from the CFM-Schiller GmbH for their support during the construction and optimization of the low vibration laboratory. Further, we would like to thank R. Temirov for fruitful discussions.

## APPENDIX

Alternatively to the spectral density representation of the vibration velocity, the one-third octave band representation is frequently used in order to present the frequency dependence of vibrations. The choice of the spectral density representation is particularly appropriate, when the disturbances are tonal and represented by peaks, as it allows the identification of the vibration sources. On the other hand for a general definition of vibration limits (where the sources of particular disturbances are not considered), manufacturers of microscopes select the one-third octave representation for the spectra, which integrates over larger frequency bands. For the same reason, the official standards for the vibration criteria of buildings are defined that way.

In order to facilitate comparability of our results to standards (or results obtained at other laboratories), we show in this appendix how to convert the spectral density representation to the one-third octave bands.

In octave band representation, the bandwidth of the bins in which the data are shown is proportional to the respective center frequency [12]. In a very common form of proportional bandwidth analysis, each bandwidth is 23% (i.e. one-third octave) of the center frequency and three adjacent bands represent an octave. The center frequencies are given by established standards (e. g. DIN EN ISO 266): starting from $f_0 = 1000$ Hz, the lower center frequencies $f_{n-1}$ is recursively defined as

$$f_{n-1} = f_n/2^{\frac{1}{3}} \approx 0.79 f_n \tag{10}$$

and the low (high) frequency limits of each band are given by $f_n^{\text{low}} = f_n/2^{\frac{1}{6}}$ and $f_n^{\text{high}} = f_n \cdot 2^{\frac{1}{6}}$, respectively.

In a first step the spectral density, which is normalized per $\sqrt{\text{Hz}}$, is referred to the very small bandwidth bins in which the spectral density spectrum is represented (in our case each bin in the spectral density is separated by $\Delta f_{\text{bin}} = 0.02$ Hz $= B$). Assuming that the spectral density within this small bin bandwidth is constant, the spectral density can be converted to what is called a narrow band spectrum. The RMS velocity in each bin is calculated by using the definition in Eq. 7.

In a second step, this narrow band spectrum is converted into the one-third octave bands by summing all velocity values of the narrow band representation inside the corresponding one-third octave band as follows

$$v_{\text{1/3 octave band at } f_c} = \sqrt{\sum_{f=0.89f_c}^{f=1.12f_c} \langle v^2 \rangle_B} \qquad (11)$$

The result for the conversion from the spectral density of the vertical direction (red curve in Fig. 6) to the one-third octave band representation is shown in Fig. 7. The quantity plotted in the one-third octave representation is the RMS velocity arising from the respective bandwidth of the one-third octave bin. This, one-third octave band representation of the vibrations is frequently used by microscope manufactures, engineers, and architects and provides comparability by the use of official standards.

---

* Corresponding author: `b.voigtlaender@fz-juelich.de`

# Erratum: "Low Vibration Laboratory with a Single-Stage Vibration Isolation for Microscopy Applications" [Rev. Sci. Instrum. 88, 023703 (2017)]

Bert Voigtländer,[1, 2, *] Peter Coenen,[1, 2] Vasily Cherepanov,[1, 2]
Peter Borgens,[1, 2] Thomas Duden,[3] and F. Stefan Tautz[1, 2]

[1]*Peter Grünberg Institut (PGI-3), Forschungszentrum Jülich, 52425 Jülich, Germany*
[2]*Jülich Aachen Research Alliance (JARA) Fundamentals of Future Information Technology, 52425 Jülich, Germany*
[3]*Konstruktionsbüro Duden, Borgsenallee 35, 33649 Bielefeld, Germany*
(Dated: December 4, 2017)

PACS numbers:

In our publication entitled "Low vibration laboratory with a single-stage vibration isolation for microscopy applications" [1] we have erroneously labeled the measured *spectrum* $N_{spec}$ of the vibrations as the *spectral density* $N_{PSD} \equiv N_v$ of the vibrations in our laboratory. Due to this the measured vibration levels (spectral densities) shown in Figs. 3, 5, 6, and 7 have to be multiplied by a factor of 5.9. The corrected figures are shown below. Note that the magenta curve in Fig. 3 of [1] is not shifted, because it was measured with a bandwidth of 1 Hz.

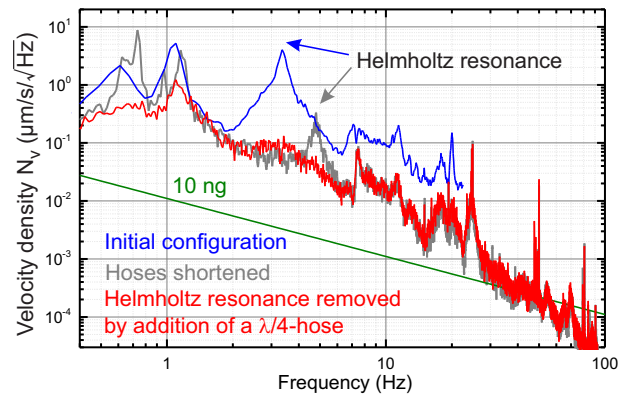

FIG. 2: (color online) Corresponds to Fig. 5 in [1]. Spectral density of the rms vibration velocity (vertical component) for different stages of the optimization of the vibration isolation. The initial state (blue) is characterized by a large peak at about 3 Hz due to the Helmholtz resonance. After shortening of the hoses the Helmholtz resonance shifted to about 5 Hz (gray). The Helmholtz resonance has been removed by destructive interference of standing waves using an additional $\lambda/4$-hose (red).

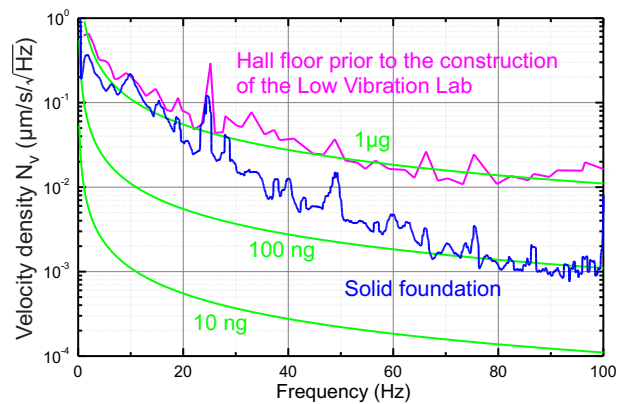

FIG. 1: (color online) Corresponds to Fig. 3 in [1]. Spectral density of the rms vibration velocity (vertical component) $N_v(f)$ on the floor of the building before any construction has been started (magenta line), compared to the rms vibration level on the solid 200 ton solid foundation (blue line). For comparison $N_v(f)$ curves (rms) corresponding to different fractions of the gravitational acceleration $g$ are shown.

While experts in spectral analysis will be aware of the difference between a *spectrum* and the *spectral density*, scientists using spectral analysis only occasionally may not be aware of the difference between both. Thus we discuss in the following this difference, and how the two can be converted into each other. Moreover, we discuss in which circumstances the one or the other should be used, and how the calibration of a spectrum or a spectral density is experimentally verified.

Nowadays, when spectrum analyzer instruments (with all the calibration steps already included) are less frequently used in favor of analogue to digital conversion of the measured signal followed by a subsequent software discrete Fourier transform (DFT), the correct calibration is no more "included" by the spectrum analyzer hardware. Since the discrete Fourier transform just transforms $n$ numbers to $n$ new numbers, the user has to take care about the necessary calibration steps, and errors may occur in this calibration. While this is straightforward in principle, it involves a number of non-trivial details. Here, we also include the description of an experimental calibration procedure which gives an easy cross-check for the correct calibration of the spectrum or spectral density. We realized that the above mentioned information is not easily found in the literature and thus this information presented compactly here.

If a continuous signal $S(t)$ is sampled with a sampling frequency $f_{sample}$, this signal is represented as a discrete time series $S(k/f_{sample})$. The discrete Fourier transform

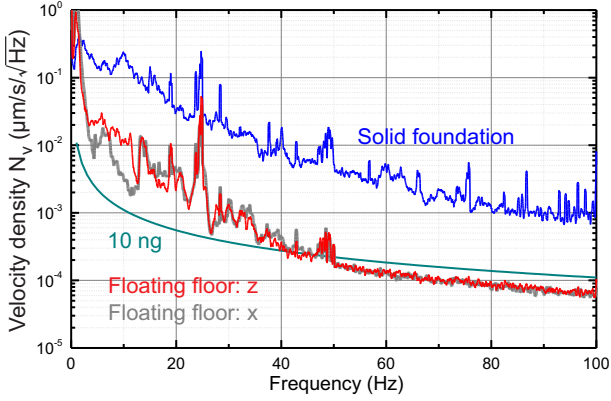

FIG. 3: (color online) Corresponds to Fig. 6 in [1]. Spectral density of the velocity (rms) measured on the floating floor in the final configuration. The vertical direction is shown in red, while one horizontal direction is shown in gray. The 10 ng line is shown in green.

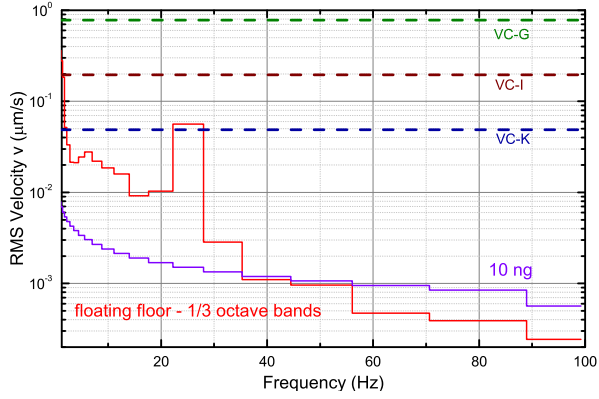

FIG. 4: (color online) Corresponds to Fig. 7 in [1]. One-third octave band analysis of the RMS velocity $v$ in $z$-direction measured on the floating floor in the final configuration. The red curve is calculated from the corresponding curve in Fig. 3. Standards for vibration isolation of buildings (vibration criteria, VC) are given for comparison (following IEST-RP-CC024), as well as the 10 ng line (in terms of one-third octave bands).

(DFT) of a time series of length $n$ is defined as

$$\hat{S}(m) = \sum_{k=0}^{n-1} S(k/f_{\text{sample}}) e^{-2\pi i k m/n},\qquad(1)$$

with $m = 0...n-1$. The power spectral density (termed PSD, or $N_{\text{PSD}}^2$) is proportional to the absolute square of the discrete Fourier transform (DFT) [2]. If we do not consider windowing yet (i.e consider a rectangular window) [3, 4], the single sided PSD results as

$$N_{\text{PSD}}^2(m) = \frac{2}{f_{\text{sample}}\, n} \left|\hat{S}(m)\right|^2,\quad m = 0...n/2. \qquad(2)$$

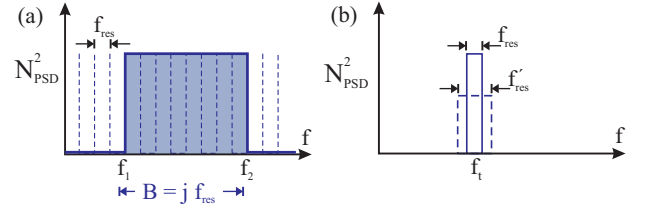

FIG. 5: (color online) (a) Case of a constant power spectral density within $B$ (blue shaded area). The DFT representation of the power spectral density has $j$ (same) values with a frequency bin with of $f_{\text{res}} = f_{\text{sample}}/n$. In this case the power spectral density is independent of the width of the frequency bin of the DFT, $f_{\text{res}}$ (c.f. Eq. (6)), while the power spectrum depends on the bin width (c.f. Eq. (10)). (b) Power spectral density of a tonal signal (sinusoidal), which has a non vanishing value only in one frequency bin. In this case DFT representation of the power spectral density depends on the frequency bin with $f_{\text{res}}$ (c.f. Eq. (7)), while the power spectrum is independent of the bin width (c.f. Eq. (9)).

Note that other definitions of the DFT than the one in Eq. 1 result in other factors in Eq. 2 [3]. The spectral density $N_{\text{PSD}}$ is the square root of the power spectral density $N_{\text{PSD}}^2$.

For a continuous signal the power spectral density of a signal is related, via Parseval's identity, to the root mean square (RMS) $S_{\text{RMS}}$ of the signal as

$$S_{\text{RMS}}^2 = \lim_{T\to\infty} \frac{1}{T}\int_0^T S^2(t)\, dt \equiv \langle S^2(t)\rangle = \int_0^\infty N_{\text{PSD}}^2(f)\, df. \qquad(3)$$

If the signal is a discrete time series, the continuous quantities $S(t)$ and $N_{\text{PSD}}(f)$ translate to discrete values as

$$S(t) \leftrightarrow S(k/f_{\text{sample}}) \text{ and } N_{\text{PSD}}(f) \leftrightarrow N_{\text{PSD}}(m\, f_{\text{res}}),\qquad(4)$$

respectively with $k = 0...n-1$, and $m = 0...n/2$. The width of the $n$ frequency bins of the DFT is given by $f_{\text{res}} = f_{\text{sample}}/n$ [3]. For a discrete signal Eq. (3) translates to

$$\begin{aligned} S_{\text{RMS}}^2 &= \frac{1}{T}\sum_{k=0}^{n-1} S^2(k/f_{\text{sample}})/f_{\text{sample}} \\ &= \sum_{m=0}^{n/2} N_{\text{PSD}}^2(m\, f_{\text{res}})\, f_{\text{res}}. \end{aligned} \qquad(5)$$

In the following we consider two simple examples for the power spectral density, a constant power spectral density (Fig. 5(a)) and a power spectral density of a tonal sinusoidal signal (Fig. 5(b)).

*Spectral density.* If the power spectral density of the signal is considered within a certain frequency bandwidth $B = f_2 - f_1$ between $f_1$ and $f_2$ (as indicated by the blue shaded area in Fig. 5(a)), the power spectral density is zero outside the range of the bandwidth $B$. We assume further that $f_\text{res} \ll B$, which is usually the case. If the power spectral density is constant for the $j$ bins between $f_1$ and $f_2$, the $N_\text{PSD}^2$ in Eq. (5) can be written in front of the sum and the sum yields $j \cdot f_\text{res} = B$. Thus for a constant PSD Eq. (5) results in

$$N_\text{PSD}^2 = \frac{S_\text{RMS}^2}{B}. \qquad (6)$$

If for example the signal is a voltage, e.g. of RMS amplitude $S_\text{RMS} = 1\,\text{V}$, and $B = 100\,\text{Hz}$, a spectral density of $N_\text{PSD} = 0.1\,\text{V}/\sqrt{\text{Hz}}$ results. In this case the (power) spectral density is independent of the width of the frequency bins. This case is desirable as the value of the (power) spectral density has a significance independent of the width of the frequency bins, i.e. independent of details of the sampling.

For the case of a tonal (sinusoidal) signal the situation is different. For the sake of simplicity we consider cases without spectral leakage present [4]. Then the tonal signal is usually located within a single non-zero frequency bin of width $f_\text{res}$ at a frequency $f_t$, as shown in Fig. 5(b). In this case only one term of the sum in Eq. (5) survives and the (power) spectral density of this bin depends (undesirably) on the width of the frequency bins $f_\text{res}$, as

$$N_\text{PSD}^2(f_\text{res}) = \frac{S_\text{RMS}^2}{f_\text{res}}. \qquad (7)$$

This means that for instance a tonal signal with an RMS amplitude of $1\,\text{V}$ results in different values for the (power) spectral density, depending on the width of the frequency bins $f_\text{res}$, as also shown in Fig. 5(b) for two different values of the frequency bin width $f_\text{res}$ and $f'_\text{res}$, respectively. This dependence of the value of the (power) spectral density on the frequency bin width $f_\text{res}$, which depends on the particular length of the time series used for the DFT and the particular sampling rate, is of course undesirable. Thus the value of the (power) spectral density for a tonal signal has no unique significance without the knowledge of some details on the sampling process, such as the sampling rate $f_\text{sample}$ and the length $n$ of the DFT.

*Spectrum.* A different quantity, the power spectrum $N_\text{spec}^2$, or the spectrum $N_\text{spec}$, defined as

$$N_\text{spec}^2 \equiv N_\text{PSD}^2 \cdot f_\text{res}, \qquad (8)$$

avoids this disadvantage. When inserting Eq. (7), valid for a tonal signal, into Eq. (8), the (power) spectrum of a tonal signal turns out to be independent of the width of the frequency bin, as

$$N_\text{spec}^2 = S_\text{RMS}^2. \qquad (9)$$

As Eq. (9) shows, the value of the spectrum is equal to the RMS amplitude of the tonal (sinusoidal) signal, $N_\text{spec} = S_\text{RMS}$ (e.g. $1\,\text{V}$).

However, undesirably for a signal of constant (power) spectral density, the spectrum $N_\text{spec}$ depends on the width of the frequency bin $f_\text{res}$, as evident when inserting Eq. (6) into Eq. (8), resulting in

$$N_\text{spec}^2 = \frac{S_\text{RMS}^2}{B} f_\text{res}. \qquad (10)$$

In conclusion, neither the (power) spectral density, nor the (power) spectrum deliver a value which is independent of the frequency bin for a tonal signal *as well as* for signal with constant PSD (representative of a broad band signal with a relatively flat PSD). A solution of this dilemma would be to choose the width of the frequency bin $f_\text{res} = 1\,\text{Hz}$, so that both, the spectral density and the spectrum have the same numeric value (but still different units, e.g. $\text{V}/\sqrt{\text{Hz}}$ and V, respectively). However, if low frequencies approaching $1\,\text{Hz}$ and below are of interest, a frequency bin with of $1\,\text{Hz}$ is too wide.

So far we have not considered the windowing in the DFT [4], which means we have so far implicitly considered a rectangular window function. When applying other window functions in the DFT, a quantity named "normalized equivalent noise bandwidth" (NENBW) can be defined and Eq. (8) is extended with $f_\text{res}^\text{eff}$ to

$$N_\text{spec}^2 \equiv N_\text{PSD}^2 \cdot f_\text{res}^\text{eff} = N_\text{PSD}^2 \cdot f_\text{res} \cdot \text{NENBW}, \qquad (11)$$

and values of NENBW for different windows are shown in Tab. (I) [3]. In order to present the complete information of a spectral analysis, both the (power) spectral density, as well as the (power) spectrum have to be presented, or one of them and $f_\text{res}^\text{eff}$.

TABLE I: "Normalized equivalent noise bandwidth" (NENBW) for different windows [3]

|  | Hanning | Flat top HP | Welch |
|---|---|---|---|
| NENBW | 1.5 | 3.4279 | 1.2 |

In the signal processing from the time series of the signal to the spectral density or spectrum several proportionality factors are involved due to the use of, for example, either RMS amplitude or peak amplitude, either two-sided spectrum or single-sided spectrum, either natural frequency PSD or angular frequency PSD, or due to different window types, etc. So one has to consider all these factors carefully. Complementary also an experimental calibration of spectral density or spectrum is very desirable and will be considered in the following.

A tonal signal e.g. from a signal generator can be used to calibrate the spectrum or the spectral density. According to Eq. (9) the RMS signal amplitude of a tonal signal corresponds directly to the amplitude of the spectrum $N_\text{spec}$, independent of the width of a frequency bin. For the calibration of the power spectral density using a tonal signal, the effective width of a frequency bin enters. According to Eq. (7) (extended to $f_\text{res}^\text{eff}$), the RMS signal amplitude of a tonal signal $S_\text{RMS}^2$ has to be divided by $f_\text{res}^\text{eff}$ in order to obtain the power spectral density.

Alternatively to a calibration with a tonal signal a signal of constant power spectral density (white noise) and known amplitude can be used for the calibration. Such a signal is provided for instance by the Johnson-Nyquist noise (thermal noise) of a resistor $R$ as as voltage source of known RMS voltage

$$U_{JN}^{RMS} = \sqrt{4\,k_B\,T\,R\,B}, \qquad (12)$$

and constant PSD (white noise spectrum). The the bandwidth $B$ in Eq. (12) corresponds to the effective width of a frequency bin of the DFT as $B = f_{res}^{eff}$. In order to obtain a reasonably large voltage, a resistor of large resistance should be used ($R \geq 500\,k\Omega$) and considered in parallel with the input resistance of the measurement device.

In conclusion, if a case of a spectral analysis includes tonal peaks, as well as (locally) constant regions as function of frequency, both the spectrum, and the spectral density are required in order to deliver quantitative results for tonal peaks and broad band regions. The tonal peaks are represented quantitatively in the spectrum (e.g. in volts), while constant regions are represented quantitatively in the spectral density (e.g. as V/$\sqrt{Hz}$). Spectrum and spectral density can be converted into each other by the proportionality factor $\sqrt{f_{res}^{eff}} = \sqrt{(f_{sample}/n) \cdot NENBW}$.

## ACKNOWLEDGMENTS

We would like to acknowledge discussions with Alex Khajetoorians and coworkers. We would like to thank Joachim Krause for critically reading the manuscript.

---